\documentclass[conference,compsoc,11pt]{IEEEtran}

\ifCLASSOPTIONcompsoc
  \usepackage[nocompress]{cite}
\else
  \usepackage{cite}
\fi

\usepackage{graphicx}
\usepackage{stfloats}
\usepackage{subcaption}
\usepackage{float}
\usepackage{enumitem}
\usepackage{amsmath}
\usepackage[hyphens]{url}

\ifCLASSINFOpdf
\else
\fi

\begin{document}
\title{Towards Carbon-Aware Container Orchestration: Predicting Workload Energy Consumption with Federated Learning}
 
\author{
\IEEEauthorblockN{Zainab Saad\IEEEauthorrefmark{1},
Jialin Yang\IEEEauthorrefmark{1},
Henry Leung\IEEEauthorrefmark{1} and
Steve Drew\IEEEauthorrefmark{1}
}
\IEEEauthorblockA{\IEEEauthorrefmark{1}Department of Electrical and Software Engineering\\
University of Calgary,
Calgary, Alberta T2N 1N4, Canada\\ Email: 
\{zainab.saad1,jialin.yang,leungh,steve.drew\}@ucalgary.ca
}}

\maketitle

\begin{abstract}
The growing reliance on large-scale data centers to run resource-intensive workloads has significantly increased the global carbon footprint, underscoring the need for sustainable computing solutions. While container orchestration platforms like Kubernetes help optimize workload scheduling to reduce carbon emissions, existing methods often depend on centralized machine learning models that raise privacy concerns and struggle to generalize across diverse environments. In this paper, we propose a federated learning approach for energy consumption prediction that preserves data privacy by keeping sensitive operational data within individual enterprises. By extending the Kubernetes Efficient Power Level Exporter (Kepler), our framework trains XGBoost models collaboratively across distributed clients using Flower’s FedXgbBagging aggregation using a bagging strategy, eliminating the need for centralized data sharing. Experimental results on the SPECPower benchmark dataset show that our FL-based approach achieves 11.7\% lower Mean Absolute Error compared to a centralized baseline. This work addresses the unresolved trade-off between data privacy and energy prediction efficiency in prior systems such as Kepler and CASPER and offers enterprises a viable pathway toward sustainable cloud computing without compromising operational privacy.
\end{abstract}

\IEEEpeerreviewmaketitle

\section{Introduction}
The increasing reliance of enterprises on large-scale data centers to run resource-intensive workloads has led to a significant rise in carbon footprint \cite{gupta2021chasing}, now accounting for approximately 2\% of global greenhouse gas emissions. As the demand for cloud services continues to grow, research and development efforts have focused on extending container orchestration tools, such as Kubernetes \cite{james2019low}, to optimize workload scheduling based on carbon emissions, thereby reducing the overall carbon footprint of data centers. Machine learning offers the potential to improve scheduling decisions by predicting energy consumption in data centers. However, centralized machine learning raises privacy concerns, as organizations prefer not to share their data with third-party servers. Moreover, models trained in isolation within individual data centers may lack the generalizability needed to make accurate predictions across diverse environments, due to their limited exposure to varied datasets. 

To address these challenges, we propose using a decentralized machine learning paradigm known as federated learning (FL). In FL, data remains within the enterprise’s servers while participating in the collaborative training of a robust global model. This decentralized model allows for the preservation of data privacy while still benefiting from the diversity of datasets, resulting in a more robust and generalizable predictive model. Using federated learning, this paper aims to contribute to the broader goal of sustainable computing, offering a viable pathway for enterprises to optimize their operations in an environmentally responsible manner.

We aim to achieve our sustainable computing goal by developing an open-source, federated learning framework with carbon awareness that can be integrated into Kubernetes, enabling carbon-aware container orchestration. The proposed solution can be deployed in cloud environments, offering enterprises a sustainable approach to reducing their carbon footprint while maintaining data privacy. We developed the federated learning framework by extending Kubernetes Efficient Power Level Exporter (Kepler). 

Previous studies have explored carbon-aware scheduling and energy consumption prediction in cloud environments, yet the critical issue of ensuring data privacy during model training remains unresolved. Existing research demonstrates significant potential in utilizing carbon-aware scheduling within Kubernetes clusters to effectively reduce carbon emissions; however, these approaches typically rely on centralized training, which introduces concerns regarding data privacy. To address this gap, our work integrates federated learning with Kubernetes, enabling accurate energy consumption predictions while preserving the privacy of distributed data sources. The contribution of this paper can be summarized as follows: applying federated learning to train energy prediction models without requiring centralized data collection and addressing the unresolved challenge of balancing data privacy with accurate energy estimation in cloud environments. This approach provides a novel alternative to existing systems like Kepler and CASPER \cite{kepler, CASPER}.

It is crucial to investigate whether federated learning can provide accurate and privacy-preserving energy predictions for cloud workloads managed by Kubernetes, thus enabling carbon-aware scheduling without requiring centralized data sharing. The main research question will be addressed:

\textbf{Does FL achieve comparable energy prediction accuracy to centralized models when evaluated on real-world cloud workloads, and effectively resolve the privacy-efficiency trade-off left unresolved by prior work such as Kepler and CASPER?}

To address this research question, we evaluate both the proposed federated learning model and a centralized baseline model using identical real-world cloud workload dataset. Performance comparisons are conducted using standard metrics, including Mean Absolute Error (MAE) and Root Mean Squared Error (RMSE), to verify if FL can achieve or surpass centralized model accuracy. Additionally, our FL approach ensures client data remains local, explicitly addressing the privacy-efficiency trade-off that previous studies have not fully resolved.

\section{Related Work}

\subsection{Machine Learning For Power Modeling}
Kepler\cite{kepler} leverages eBPF to probe performance counters and other system statistics, employs machine learning models to estimate workload energy consumption based on these statistics, and exports them as Prometheus metrics. Kepler collects power metrics from system hardware such as CPU and DRAM power consumption, using RAPL and sensors like ACPI. The exporter also uses Berkeley Packet Filter (BPF) programs to collect kernel-level hardware performance events related to processes. This data is then passed to the Kepler Model Server. During the training phase, the Model Server utilizes data collected from hardware counters and real-time power metrics to develop power models. The trained model is used to estimate power consumption at the process level, taking into account factors such as CPU cycles, cache hits, and cache misses.

Kepler's robust modeling\cite{robustKepler} extension goes further by supporting accurate power estimation even in the absence of direct hardware measurements. It enables container and process-level power attribution in virtualized and multi-tenant environments by training machine learning models on a wide range of hardware counters, such as CPU instructions, context switches, cache metrics and correlating them with known power measurements collected during training. These models are hardware-agnostic and generalize across different platforms,  allowing real-time inference of power consumption in production workloads. This enables container-level energy estimation in cloud-native environments, even in virtualized or restricted systems where direct access to physical energy sensors is unavailable

Several heuristic and meta-heuristic scheduling approaches, such as genetic algorithms and ant colony optimization (ACO)\cite{ACO}, have been applied to task scheduling in cloud computing environments. While these methods can optimize for objectives like makespan and cost, they often do not explicitly incorporate energy efficiency into their optimization criteria.

Reinforcement learning-based schedulers, such as DeepRM\cite{DeepRM} and Decima\cite{Decima}, have been developed to improve scheduling efficiency. DeepRM focuses on minimizing average job slowdown and completion time without considering energy consumption. Similarly, Decima aims to minimize average job completion time but does not explicitly include energy consumption in its optimization objectives.

Smart-Kube\cite{Smart-Kube} is a DRL approach to address the sequential decision-making problem of job scheduling in Kubernetes, aiming to maintain a target utilization of resources (e.g., CPU, memory) while minimizing the number of active nodes, thereby reducing the cluster's energy consumption.

While centralized machine learning approaches can enhance scheduling by predicting data center energy consumption, they introduce significant privacy risks due to the requirement of data sharing with external servers \cite{privacyConcern}. These concerns, along with limited generalization in isolated settings, motivate the exploration of decentralized alternatives.

\subsection{Carbon-Aware Scheduling in Distributed Systems}
The rising global awareness regarding the environmental impacts of computing has spurred significant research into carbon-aware scheduling strategies. Microsoft's whitepaper, \textit{Carbon-aware computing: Measuring and reducing the carbon intensity associated with software in execution}\cite{carbon_aware_whitepaper}, emphasizes measuring software-related carbon emissions and incorporating real-time carbon intensity data into scheduling decisions to shift workloads to cleaner times and regions.

CASPER (Carbon-Aware Scheduling and Provisioning for Distributed Web Services)\cite{CASPER} further advances this concept by dynamically scheduling web services across geo-distributed data centers based on regional carbon intensity, achieving notable emission reductions without compromising performance.

However, centralized methods like CASPER and the Microsoft whitepaper often suffer from scalability challenges in complex cloud environments \cite{scaleConcern} and require sharing sensitive data, raising privacy concerns. Such limitations highlight the need for decentralized, privacy-preserving methods to achieve sustainable computing effectively.

\subsection{Federated Learning for Sustainable AI}
The environmental implications of federated learning have gained increasing attention in recent years. Qiu et al. \cite{qiu2023lookcarbonfootprintfederated} conducted a pioneering study that systematically quantified the carbon footprint of FL. Their findings revealed that, under certain configurations, FL can emit up to two orders of magnitude more carbon than centralized machine learning, primarily due to the energy-intensive communication between distributed clients and servers .

Building upon this, Green Federated Learning (Green FL)\cite{GreenFL} emphasizing the trade-offs between energy efficiency, model performance, and training time. By analyzing real-world FL deployments across millions of devices, they provided insights into optimizing FL parameters to minimize carbon emissions without compromising accuracy or convergence speed.

Carbon-Efficient Federated Learning (CEFL)\cite{CEFL} further advances this direction by introducing adaptive, cost-aware client selection strategies to minimize carbon emissions. Unlike traditional FL approaches that treat all resources uniformly and focus solely on time-to-accuracy, CEFL employs a framework that takes into account different operational costs, in particular the carbon intensity of energy consumed during local training. By prioritizing clients that provide the best utility-to-cost ratio, CEFL significantly reduces carbon emissions while maintaining model performance.

More recently, Carbon-Aware Federated Learning (CAFE)\cite{CAFE}, a framework designed to optimize FL training within a fixed carbon footprint budget. CAFE incorporates predictive models for future carbon intensity and employs strategies like coreset selection and Lyapunov optimization to balance learning performance with environmental impact.

\subsection{Privacy-Preserving FL in Cloud-Native Environments}
The default flat network model of cloud-native environments allows for unrestricted communication between pods, potentially giving malicious FL clients access to other clients' sensitive data or resources. Moreover, implementing secure aggregation protocols in dynamic cloud environments can be complex due to factors such as pod rescheduling and network changes \cite{privacyConcern}.

KubeFlower\cite{KubeFlower} is a Kubernetes-native framework designed to enhance privacy in FL deployments across cloud-edge environments. It introduces Privacy-Preserving Persistent Volume Claimer (P3-VC), which adds calibrated noise to data and manages a privacy budget. Additionally, KubeFlower emphasizes isolation-by-design, which ensures that each FL client operates in its own isolated network environment to prevent unauthorized inter-client communication. To address the complexity of secure aggregation in distributed systems, KubeFlower integrates with the Flower framework. Flower is a flexible and extensible federated learning framework that facilitates coordination, communication, and aggregation among distributed clients \cite{flower}. While KubeFlower focuses on privacy through noise injection and network isolation during FL deployments, our method uses federated XGBoost model to estimate energy use accurately. So, while both methods protect privacy, our work extends the utility of FL beyond privacy preservation by using FL for the broader purpose of sustainable and energy-efficient cloud computing.

\section{Methodology}
\begin{figure}[H]
    \centering
    \includegraphics[width=1.0\linewidth]{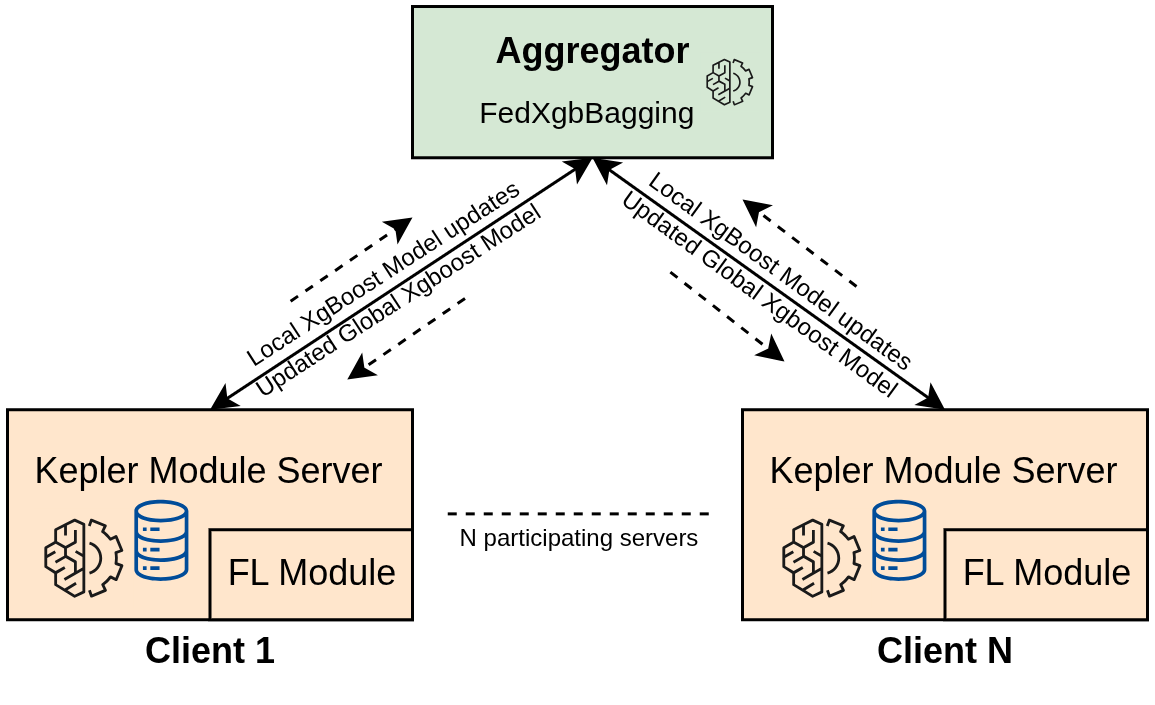}
    \caption{Federated learning architecture}
    \label{fig:kepler_fl_arch}
\end{figure}

The system architecture for training a federated learning model is shown in Figure \ref{fig:kepler_fl_arch}

\subsection{Dataset}
\subsubsection{SPECPower Benchmark Dataset}
We tested the FL implementation on the \texttt{SPECpower\_ssj2008} benchmark dataset, which is a standard for evaluating server power and performance \cite{specpower2008}. The dataset provides metrics on server utilization and energy consumption under various workload conditions. For the purpose of federated training, the dataset was synthetically partitioned to simulate a realistic multi-tenant setting where each partition represents a different organization or data center with its own local data distribution. This dataset provides:

\begin{itemize}
    \item Server power consumption under varying load levels (0\% to 100\%).
    \item Performance metrics (CPU utilization, memory usage, disk I/O).
    \item Energy efficiency ratios (performance per watt).
\end{itemize}

\subsection{Federated Learning Framework}
Our federated XGBoost model was utilized for the energy consumption prediction. The central server/aggregator initializes a global model and coordinates multiple training rounds. The clients/data centers trained local XGBoost regression models on their private datasets and sent model updates to the server. The federated XGBoost model is implemented as an extension of Kepler's power modelling pipeline.

\begin{figure}[H]
    \centering
    \includegraphics[width=1\linewidth]{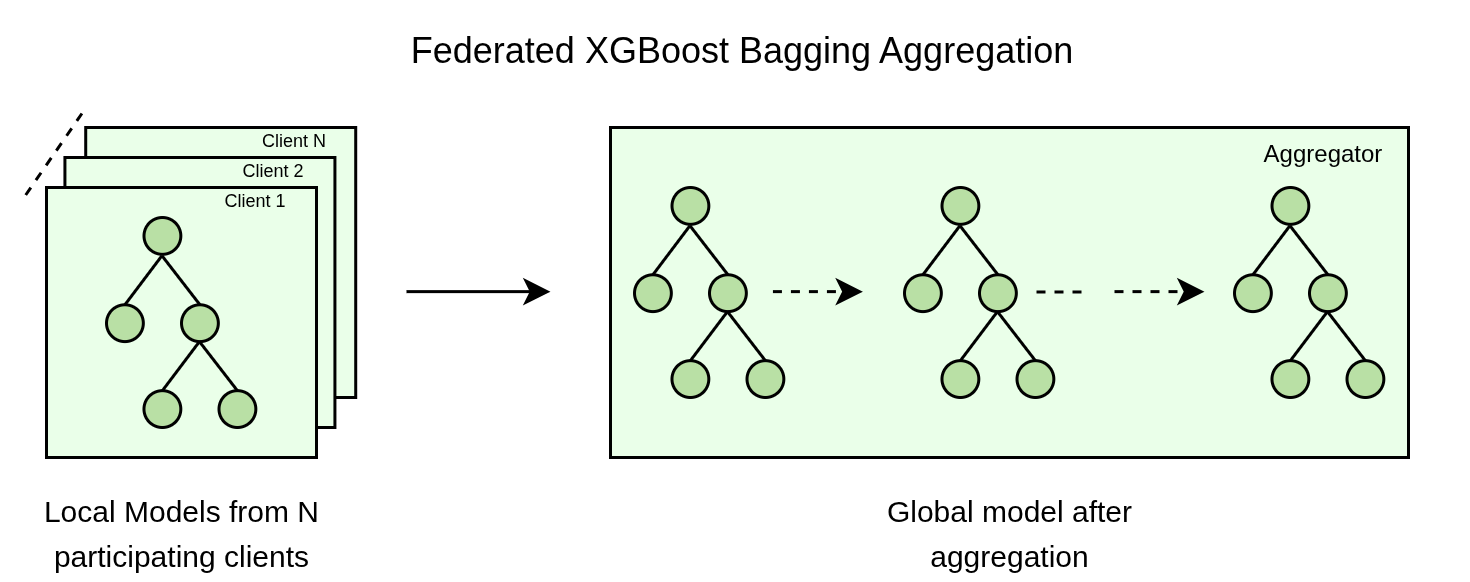}
    \caption{Federated XGBoost Bagging Aggregation Strategy}
    \label{fig:fedxgbbagging}
\end{figure}

\subsubsection{FedXgbBagging: Federated Aggregation for XGBoost Models}
The aggregation strategy used is Flower's FedXgbBagging, an ensemble-based FL aggregation strategy used to aggregate XGBoost (Extreme Gradient Boosting) models in federated learning. Unlike traditional federated averaging (FedAvg), which averages model weights (common in neural networks), FedXgbBagging aggregates decision trees from distributed clients using a bagging (bootstrap aggregating) approach as shown in Figure \ref{fig:fedxgbbagging}. Unlike neural networks, these XGBoost models cannot be averaged directly, instead these decision trees are shared (without the raw data). These models may differ slightly in tree structure or depth due to differences in local distributions, but all follow shared configuration parameters. Each participating client sends their trained trees to the central server/aggregator. Since clients may produce a different number of trees, the aggregator treats each tree independently and does not require clients to produce a fixed number of trees. This allows the global model to grow flexibly across rounds while still preserving client-specific learning. The server collects trees from all clients and constructs a global ensemble using a bagging strategy. The aggregated global XGBoost model is sent back to clients for further training or inference.
The trees from each client are exported to the server in serialized JSON form. On the server side, the aggregation function parses each client's serialized model to extract the learned decision trees. It then renumbers the trees with unique IDs to avoid collisions. These trees are appended to a shared global model structure. After this, the total number of decision trees in the global XGBoost model is updated and starting index of each training iteration's trees in the global model is marked.
To improve clarity, the mathematical formalization of FedXgbBagging can be done as follows:

Let $K$ be the number of clients and let client $k$ train an XGBoost model with $T_k$ decision trees:
\[
M_k = \{f_{k}^{(1)}, f_{k}^{(2)}, \dots, f_{k}^{(T_k)}\}, \quad \text{for } k = 1, 2, \dots, K
\]
where $f_{k}^{(t)}$ denotes the $t$-th decision tree learned by client $k$.

The server aggregates these trees to form the global ensemble $M_G$:
\[
M_G = \bigcup_{k=1}^{K} \{f_{k}^{(1)}, f_{k}^{(2)}, \dots, f_{k}^{(T_k)}\}
\]

The total number of trees in the global model is:
\[
T = \sum_{k=1}^{K} T_k
\]

During aggregation, each incoming tree $f_{k}^{(t)}$ is assigned a unique ID to avoid collisions. 

Flower's FedXgbBagging uses a uniform bagging strategy which means no weighting is applied during aggregation and each decision tree contributes equally to the final prediction.
This process is repeated in every communication round. At each round, the aggregated global model (now an ensemble of all client models) is shared back with the clients to continue training.

\subsubsection{Client Side Local Model Training}
The data pipeline loads and preprocesses SPECPower benchmark data, including CPU utilization and power metrics, for each client. The data is partitioned by node\_type to simulate heterogeneous clients, such as different server configurations. For feature engineering, the system uses BPFOnly (Berkeley Packet Filter features) as the feature group for model inputs and isolates idle power states via MinIdleIsolator to improve prediction accuracy. The current implementation relies on BPF-only features and a single energy source, ACPI, which is supported by the SPECPOWER dataset. The Isolator (MinIdleIsolator) is used to isolate relevant intervals for model training. The SpecClient, a custom Flower client class, overrides the training logic to handle node-specific energy consumption data. Each client trains the XGBoost model locally on its node-type-specific data, meaning each model is trained for specific hardware specifications. Each client trains a local XGBoost regression model on its private dataset using certain configurations. Each model consists of 100 boosting rounds (n\_estimators=100) which allows the model to iteratively correct errors from previous trees. A low learning rate of 0.01 is used to ensure smooth convergence and prevent overfitting. Instead of averaging model weights (as done in neural networks) the trained decision trees are aggregated across clients using a bagging strategy. The aggregator collects serialized decision trees from all clients and constructs a global ensemble model by combining them, forming the updated global model without accessing any raw client data.

\subsection{Privacy Protection of Data}
Our federated learning method ensures that raw data remains local to each client. During the training process, only model artifacts (decision trees in this case) are transmitted to the central aggregator. This reduces the risk of raw data leakage, in contrast to centralized machine learning approaches where data must be pooled onto a single server.

While no explicit differential privacy mechanisms are currently applied, the model’s design inherently improves data confidentiality by not exposing actual energy consumption metrics or workload characteristics directly. However, model inference attacks on shared trees remain a theoretical concern. Overall, this approach provides a baseline level of privacy by design and ensures that data locality is preserved throughout the model lifecycle.

\section{Experimental Results}
To evaluate the performance of our federated learning framework for energy prediction in Kubernetes, we conducted experiments using the SPECPower benchmark dataset. The regression model was evaluated using several metrics: Mean Absolute Error (MAE), Mean Squared Error (MSE), Root Mean Squared Error (RMSE), Mean Absolute Percentage Error (MAPE) and ($R^2$) score. The performance of the model was analyzed under varying test sizes, learning rates and training rounds, comparing client-specific and aggregated results against a centralized baseline mean absolute error pf 14.51. All experiment evaluation results were conducted with 3 clients representing heterogeneous data centers. The choice of 3 clients is because of the size of the SPECPower dataset and splitting the available data in more clients meant that there would be lesser data at each client which hindered model learning abilities. This experimental setup was designed for offline evaluation using the SPECPower benchmark dataset, which is not collected in real-time from an operational Kubernetes cluster. In future work, we plan to integrate the framework with live Kubernetes deployments across heterogeneous nodes to validate real-world scalability.

\begin{figure}[h]
    \centering
    \includegraphics[width=0.9\linewidth]{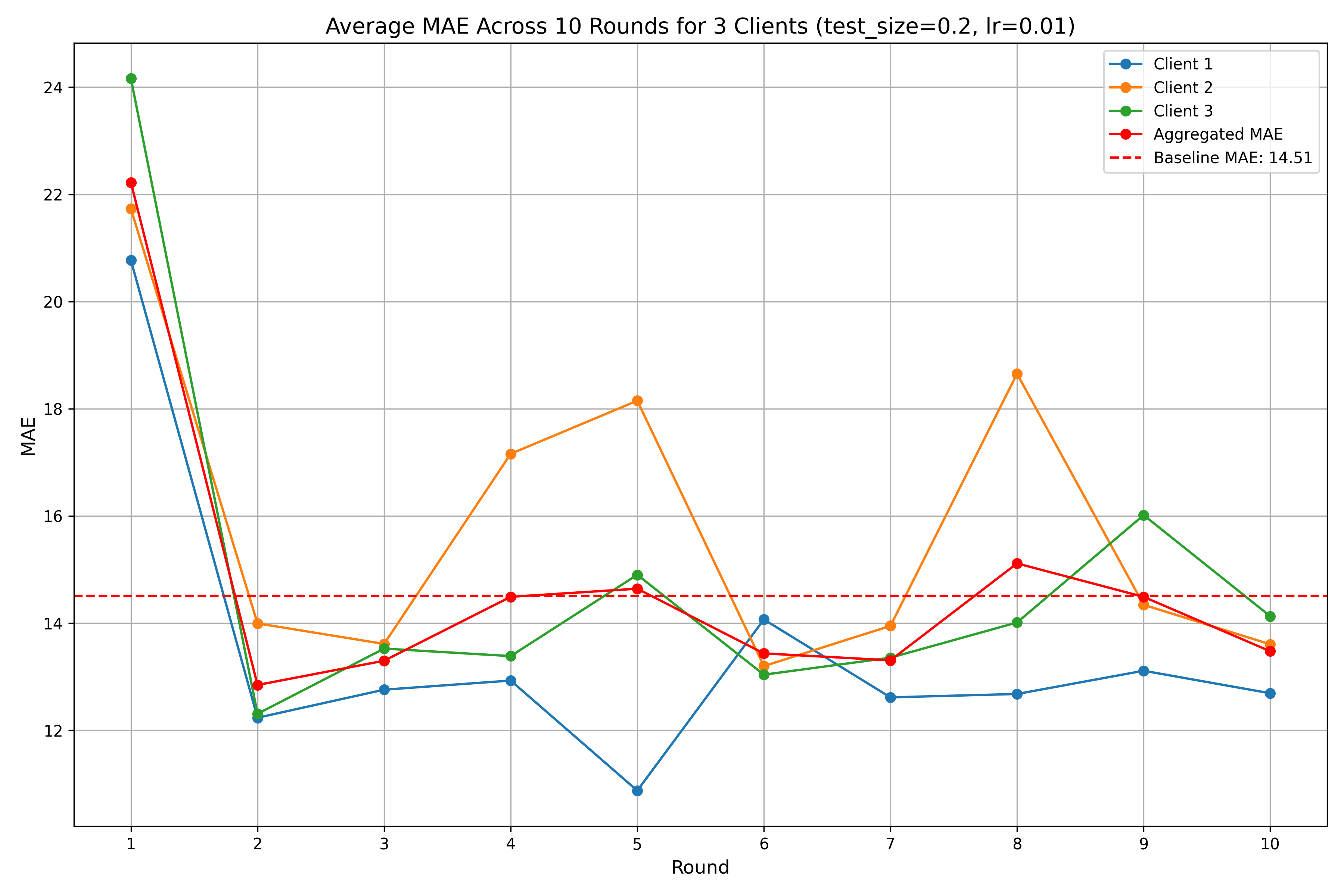}
    \caption{Mean Absolute Error Plot comparing the baseline centralized XgBoost vs the federated XgBoost Model}
    \label{fig:mae_plot}
\end{figure}

\begin{figure*}[t]
    \centering
    \begin{subfigure}{0.24\textwidth}
        \includegraphics[width=\linewidth]{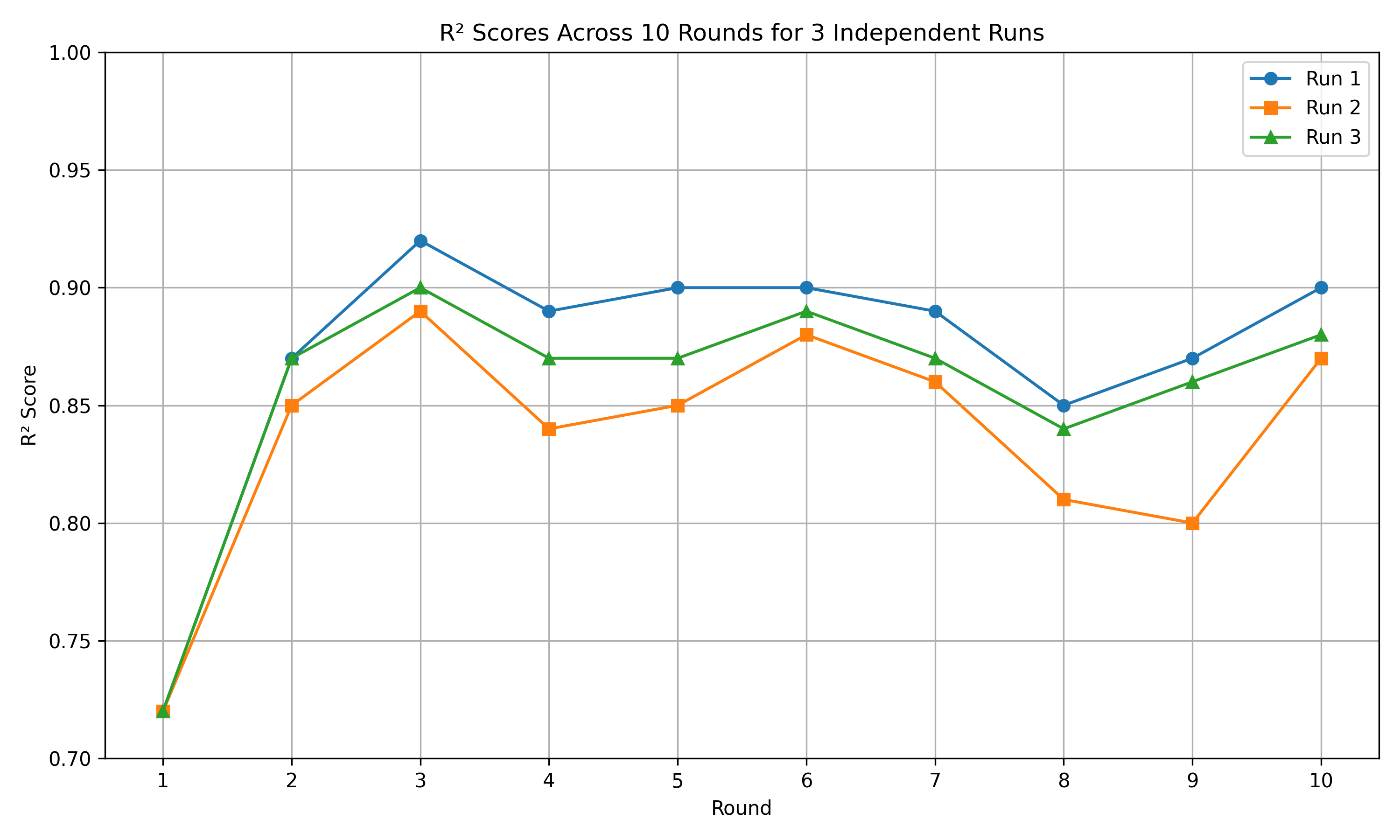}
        \caption{R-squared across 3 runs (10 rounds)}
        \label{fig:r_sq_plot}
    \end{subfigure}
    \hfill
    \begin{subfigure}{0.24\textwidth}
        \includegraphics[width=\linewidth]{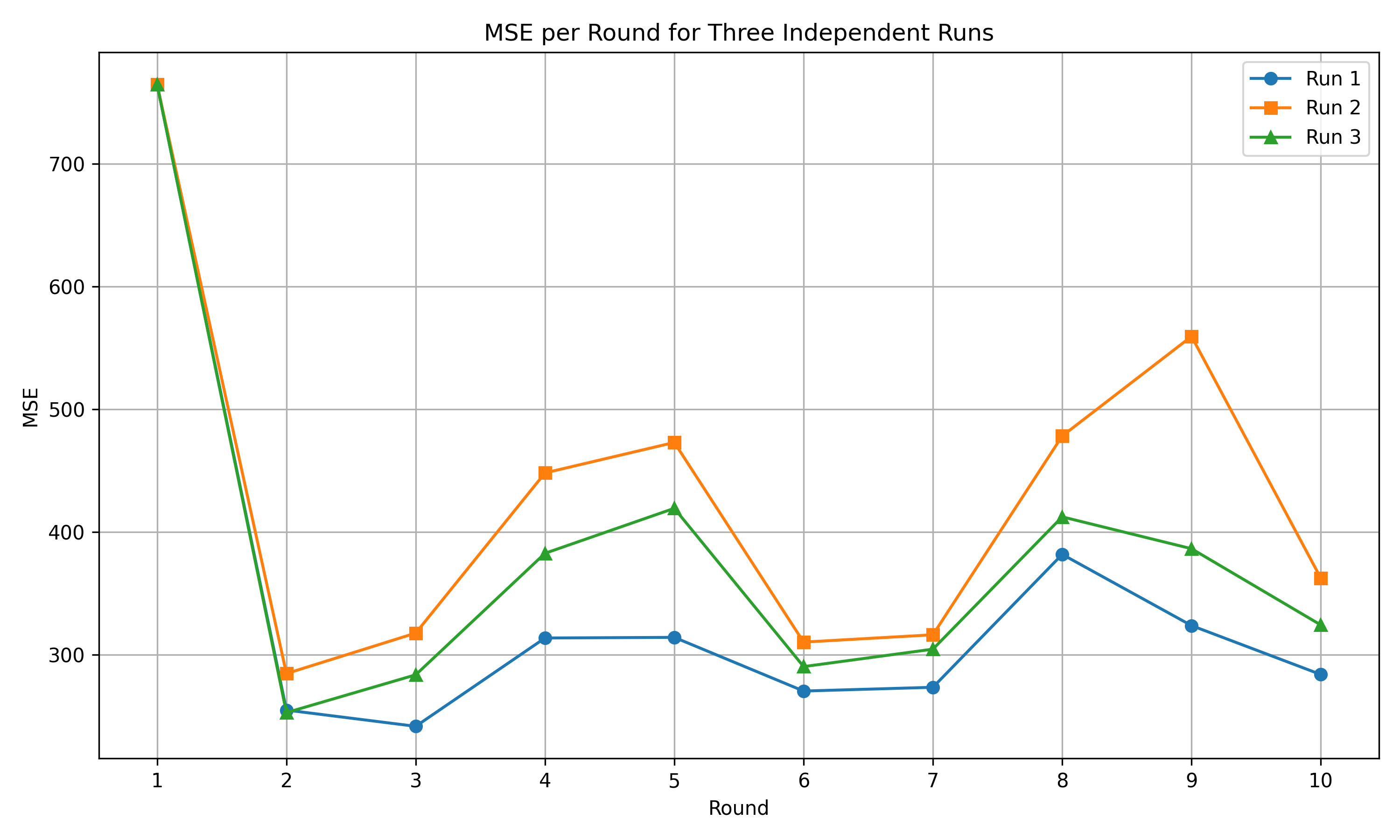}
        \caption{MSE across 3 runs (10 rounds)}
        \label{fig:mse_plot}
    \end{subfigure}
    \hfill
    \begin{subfigure}{0.24\textwidth}
        \includegraphics[width=\linewidth]{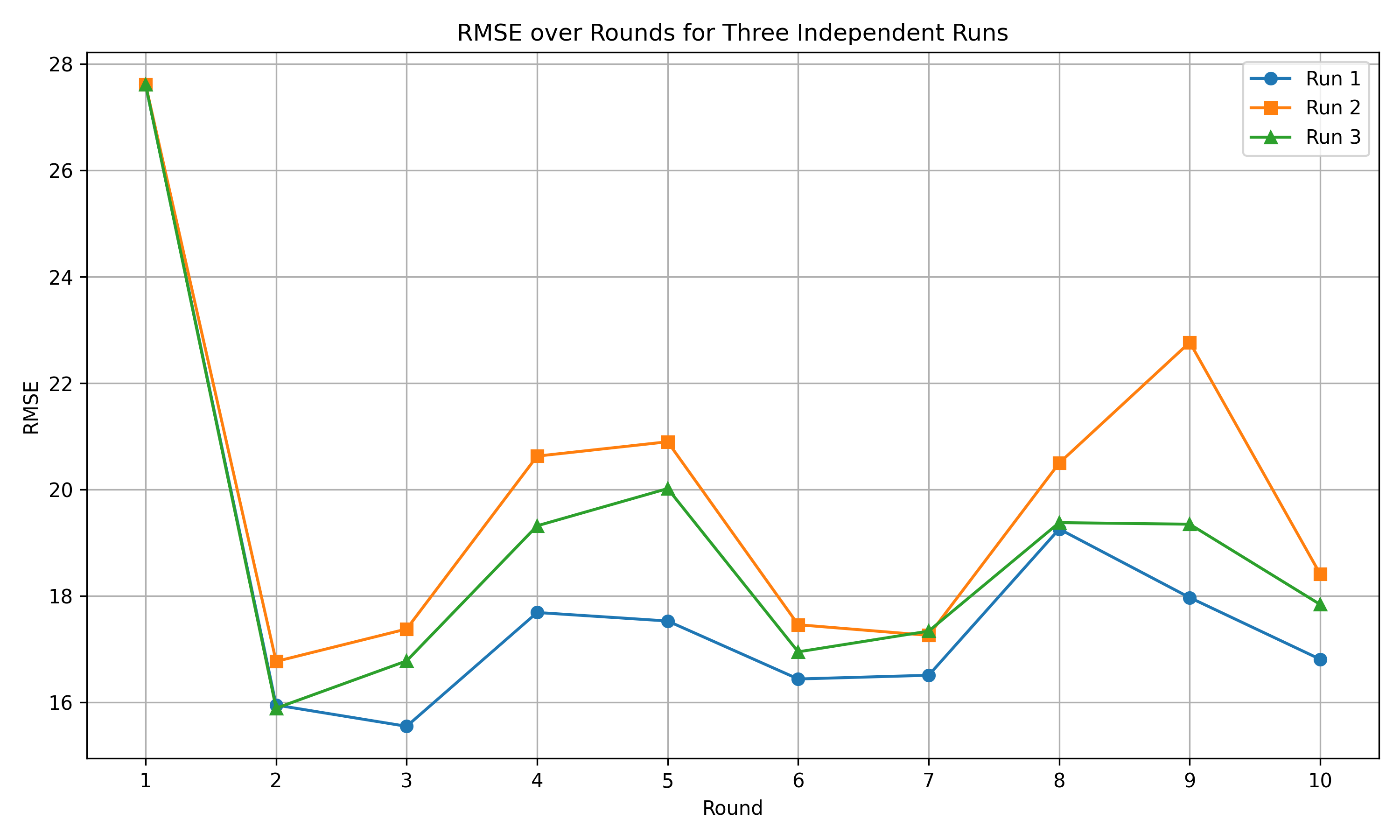}
        \caption{RMSE Error across 3 runs (10 rounds)}
        \label{fig:rmse_plot}
    \end{subfigure}
    \hfill
    \begin{subfigure}{0.24\textwidth}
        \includegraphics[width=\linewidth]{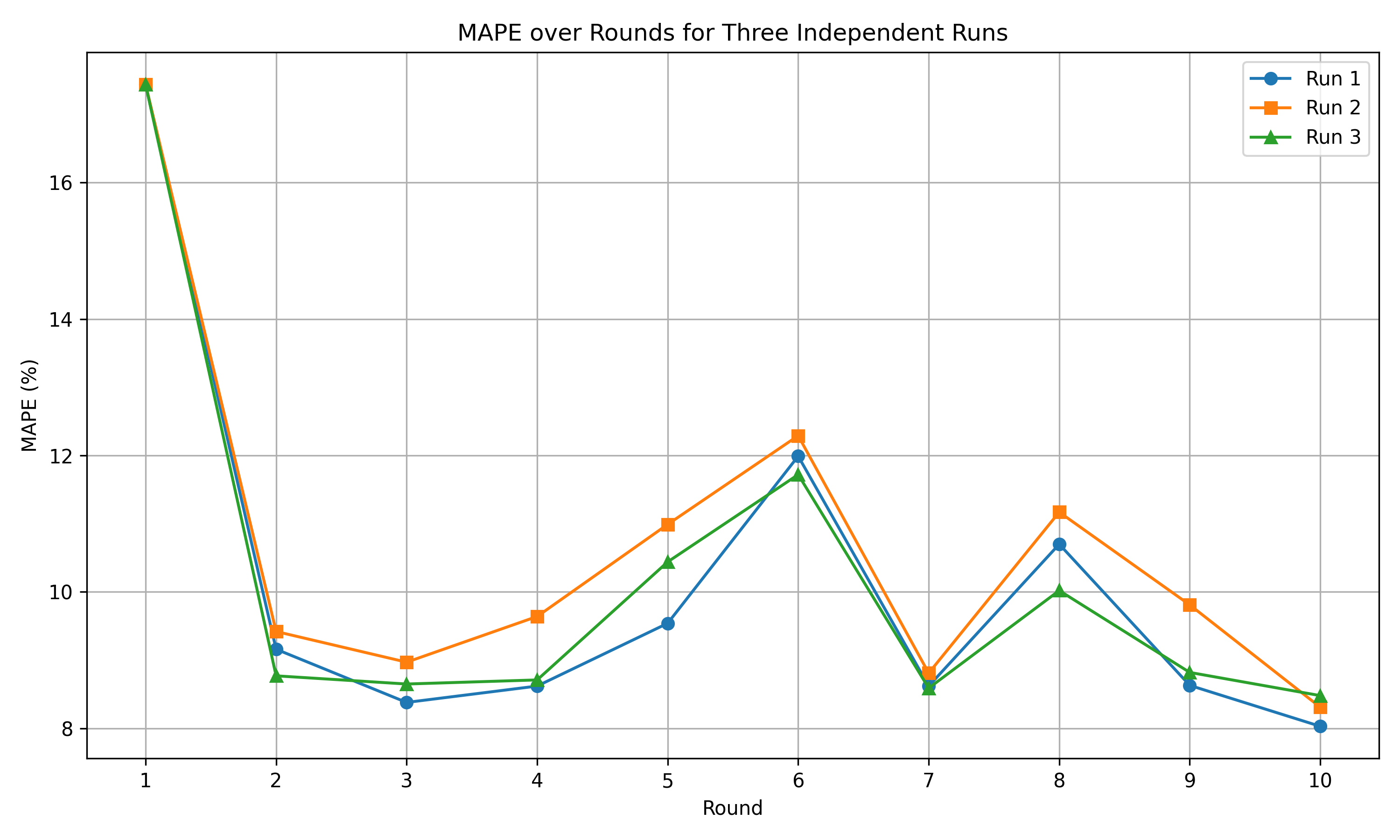}
        \caption{MAPE Error across 3 runs (10 rounds)}
        \label{fig:mape_plot}
    \end{subfigure}
    \caption{Evaluation metrics for federated XGBoost model (averaged over 3 independent runs).}
    \label{fig:eval_results}
\end{figure*}

For all experiments, SPECPower benchmark dataset was synthetically partitioned to simulate a multi-tenant environment.  Figures \ref{fig:mae_plot}, \ref{fig:r_sq_plot}, \ref{fig:mse_plot}, \ref{fig:rmse_plot}, \ref{fig:mape_plot} show the evaluation results for 20\% data reserved for testing and learning rate set to 0.01, with training done for 10 rounds. Initially, all clients start with relatively high MAE (20-24). Then, a steady improvement is observed across rounds, with the MAE decreasing from approximately 22 to approximately 12.8.

Figure \ref{fig:mae_plot} shows the MAE over 10 training rounds for three independent runs. Initially, all clients started with relatively high MAE values (approximately 20–24). As training progressed, a steady improvement was observed, with the MAE decreasing from approximately 22 to around 12.8 by the end of the 10th round. The aggregated model consistently outperformed individual clients after the initial rounds. Notably, the final aggregated MAE (12.81) is lower than the centralized baseline (14.51) by 11.7\% which shows the federated learning performance is comparable to centralized approach. 

Figure \ref{fig:r_sq_plot} shows ($R^2$) scores across 10 rounds for three independent runs. The ($R^2$)  score measures the proportion of variance in the target variable explained by the model. As training progressed, the ($R^2$) scores increased, indicating improved model fitting.

Figure \ref{fig:mse_plot} shows the MSE across 10 rounds for three independent runs. Similar to MAE, the MSE had a downward trend as training progressed. Root mean squared error and mean absolute percentage error followed a similar trend, as shown in Figure \ref{fig:rmse_plot} and \ref{fig:mape_plot} respectively.

\section{Conclusion}
This study shows that federated learning enables accurate and privacy-preserving energy consumption prediction for carbon-aware container orchestration in Kubernetes. Our FL-based framework achieved an 11.7\% lower MAE compared to centralized approach on the SPECPower benchmark while preserving data locality and addresses the privacy-efficiency trade-off in prior work. These results validate FL as a viable strategy for sustainable cloud computing and offers enterprises a pathway to reduce emissions without compromising operational data privacy. Future work will optimize FL training under dynamic carbon intensity patterns and evaluate real-world deployments in heterogeneous cluster environments.
\bibliographystyle{IEEEtran}
\bibliography{IEEEabrv,refs}
\end{document}